\documentclass[10pt,twocolumn]{article}

\usepackage{multicol}
\usepackage{graphicx}
\usepackage{amsmath}
\usepackage{lipsum}
\usepackage{color}
\usepackage{siunitx}
\usepackage{authblk}
\usepackage[margin=2cm]{geometry}

\title{Tuning the receding contact angle on hydrogels by addition of particles}
\author[1,2]{Fran\c{c}ois Boulogne}
\author[1,3]{Fran\c{c}ois Ingremeau}
\author[2]{Laurent Limat}
\author[1]{Howard A. Stone}
\affil[1]{Department of Mechanical and Aerospace Engineering, Princeton University, Princeton, NJ 08544, USA}
\affil[2]{Laboratoire Mati\`ere et Syst\`emes Complexes (MSC), UMR 7057 CNRS, Universit\'e Paris Diderot, B\^atiment Condorcet, 10 rue Alice Domon et L\'eonie Duquet, Paris, France}
\affil[3]{LIPhy, CNRS, and Universit\'e Grenoble Alpes, 140 Rue de la Physique, 38402 Saint-Martin-d'H\`eres, France}
\date{\today}
\begin{document}


\twocolumn[
    \begin{@twocolumnfalse}
        \maketitle
        \begin{abstract}
            Control of the swelling, chemical functionalization, and adhesivity of hydrogels are finding new applications in a wide range of material systems.
            We investigate experimentally the effect of adsorbed particles on hydrogels on the depinning of contact lines.
            In our experiments, a water drop containing polystyrene microspheres is deposited on a swelling hydrogel, which leads to the drop absorption and particle deposition.
            Two regimes are observed: a decreasing drop height with a pinned contact line followed by a receding contact line.
            We show that increasing the particles concentration increases the duration of the first regime and significantly decreases the total absorption time.
            The adsorbed particles increase the pinning force at the contact line.
            Finally, we develop a method to measure the receding contact angle with the consideration of the hydrogel swelling.
        \end{abstract}
    \end{@twocolumnfalse}
]

\section{Introduction}

Hydrogels are promising materials for numerous applications such as superabsorbant materials, soft contact lenses \cite{Holly1975,Ketelson2005}, drug delivery systems \cite{Hoare2008}, human implants or tissue engineering \cite{Hunt2014}.
The control of interfacial properties represents a major breakthrough in technological innovations with a particular focus on the control of roughness, adhesion \cite{Rose2014}, friction \cite{Stoimenov2013}, wetting \cite{Ketelson2005} or optical properties.
However, it is challenging to control these interfacial characteristics due to the complexity of the physical and chemical properties of hydrogels.
Among the possibilities to control these properties, some techniques consist in grafting polymers at the gel surface so as to control adhesion properties \cite{Molina2012,Sudre2012} or in patterning the surface to tune the adhesion \cite{Poulard2011}.

In the literature on colloidal deposition, it is well-known that particles can pin a contact line \cite{Deegan2000,Rio2006,Bodiguel2010,Weon2013}.
Thus, synthetic textured surfaces have been designed to achieve control of the roughness and to study contact angle hysteresis \cite{Ramos2003,Quere2008a,Forsberg2010,Paxson2013}.
The effect of defects on a receding contact line has been studied theoretically \cite{Joanny1984,Pomeau1985,Rosso2002,Moulinet2004,Doussal2009,LeDoussal2010} and observed experimentally \cite{Meglio1990,Meglio1992,Nadkarni1992,Decker1997,Reyssat2009,Delmas2011};
see \cite{DeGennes1985,Bonn2009} for reviews.

Surprisingly, despite the fact that hydrogels contain a large quantity of water, a water drop deposited on its surface has a non-zero contact angle.
This effect is due to the free polymer chains, which are present at the gel interface and pin the contact line \cite{CohenStuart2006}.
Therefore, to tune the wetting properties of hydrogels, we propose to investigate the functionalization of gel interfaces by the deposition of particles.

Previously, we have studied hydrogels, which swell following the absorption of a water drop containing a dilute suspension of micron-size particles \cite{Boulogne2015b}.
Unlike the so-called coffee stain effect observed during the evaporation of such drops on non-porous substrate, we have shown that the absorption leads to a nearly uniform deposition of particles.
In our earlier study, we highlighted that the final deposition is the result of a combination of two successive regimes: a pinned contact line followed by a receding contact line as the absorption proceeds.
We argued and observed that solvent absorption into the hydrogel occurs with a nearly homogeneous flux that decays as $t^{-1/2}$ for most of the absorbing surface.
The larger flux observed near the contact line brings more particles to the drop edge during the pinned contact line regime.
However, the second regime of the time dependent dynamics populates the center of the wetting area more than the edge, which finally erases the larger concentration at the drop edge.

In this paper, we show that a higher particle concentration leads to a stronger pinning of the contact line, \textit{i.e.} a smaller depinning contact angle.
As a result, it delays the transition between the two regimes and it affects the final deposition morphology.
Moreover, we quantify this effect, by measuring the liquid-gel contact angle with a novel method based on the detection of the hydrogel interface on which particles are adsorbed.
We rationalize our findings with the literature related to  the  pinning  of a contact line on surfaces containing defects.

\section{Experimental details}

\begin{figure}
    \centering
    \includegraphics{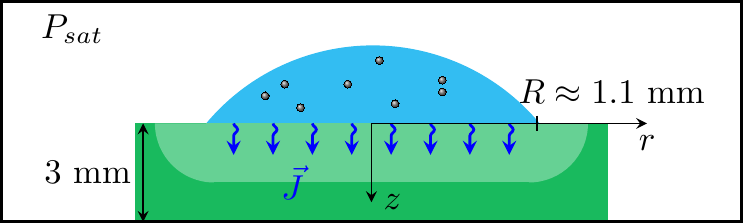}
    \caption{Experimental setup. A drop is dispensed on an absorbing hydrogel in an environment saturated in vapor to prevent evaporation.
        The light green area shows the typical swelling front \cite{Boulogne2015b}.
    }\label{fig:setup}
\end{figure}

Our experiments consist in a colloidal drop containing micron-size particles, dispensed on a swelling hydrogel (figure~\ref{fig:setup}).
We made a fully closed glass cell constituted of two concentric compartments.
The gel is placed in the center and is surrounded by a reservoir of water.
Therefore, the environment is maintained at a relative humidity of 100\% to prevent evaporation of the drop and the hydrogel.

\subsection{Hydrogels}

Absorbing substrates are made of polydimethyl acrylamide (PDMA) hydrogel \cite{Sudre2011}.
We denote $m_{mono}$ the mass of monomer (N,N-dimethylacrylamide) and $m_{w}$ the mass of water.
In addition, $[\rm{MBA}]$ and $[\rm{mono}]$ are, respectively, molar concentrations of the crosslinker (N,N'-methylene-bis-acrylamide) and the monomer.
The composition of the gel is set by $m_{mono} / (m_{mono} + m_{w}) = 0.15$ and $[\rm{MBA}]/[\rm{mono}] = 2$\%.
The quantity of initiator (potassium persulfate and N,N,N',N'-tetramethylethylenediamine) is set to a molar ratio of 0.01 of the monomer quantity.
All chemicals are purchased from Sigma-Aldrich, USA.
The solution is poured in a mold made of two glass plates ($75\times50$ cm$^2$, Dow Corning) separated with a rubber spacer of $3$ mm thickness (McMaster-Carr).
Hydrogels are stored in a vapor saturated environment and they are used three days after preparation.

\subsection{Colloidal drop}

Colloidal suspensions consist of hydrophilic fluorescent particles of $2a=1$ $\mu$m diameter (Lifetechnologies, Orange with absorption/emission wavelengths at 540/560 nm) purchased as an aqueous suspension containing 0.02 \% thimerosal.
The suspension is diluted with deionized water at a concentrations $C_p$ from $2\times 10^{6}$ to $1\times10^{9}$ particles/ml, which corresponds to a volume fraction range of $[9\times 10^{-6},4\times10^{-3}]$.
In all experiments reported here, the drop volume is $0.8$ $\mu$l.
For complementary investigations (see the appendix), we used hydrophilic fluorescent particles of $2a=0.2$ $\mu$m diameter (Lifetechnologies, Orange (540/560 nm), 2 mM azide).
Blended suspensions of non-fluorescent and fluorescent particles were prepared by mixing polystyrene-carboxylated particles of $2a=1$ $\mu$m diameter (Polysciences, Polybead carboxylate) with fluorescent particles of $2a=1$ $\mu$m diameter.
The concentration of fluorescent particles is set to $5\times 10^6$ particles/ml and the non-fluorescent particle concentrations is adjusted to obtain a given final concentration.

\subsection{Optical microscopy}

Bright field and fluorescent images are captured with a Hamamatsu camera (Digital camera ORCA-Flash4.0 C11440, $2048\times 2048$ pixels) mounted on an inverted microscope (Leica DMI4000 B) with a $5\times$ objective.
The acquisition is automated with the software Micromanager \cite{Edelstein2010}.

To measure the gel interface position, we use a confocal microscope Leica TCS SP5 with a dry objective PL Fluotar, 10$\times$.
The microscope is equipped with a translation stage and a $z$-piezo stage.

\section{Observations}

\subsection{Absorption regimes}

For a drop deposited on an absorbing hydrogel, two regimes can be distinguished, independently of the presence or absence of particles \cite{Kajiya2011}.
During regime I, the contact line is pinned, the drop has a constant radius $R$ and the drop height decreases in time.
At the end of this regime, at $t=t_{I}$, the contact line recedes (regime II) toward the center of the wetting area until the drop vanishes at time $t=t_f$.

\begin{figure}
    \centering
    \includegraphics[width=\linewidth]{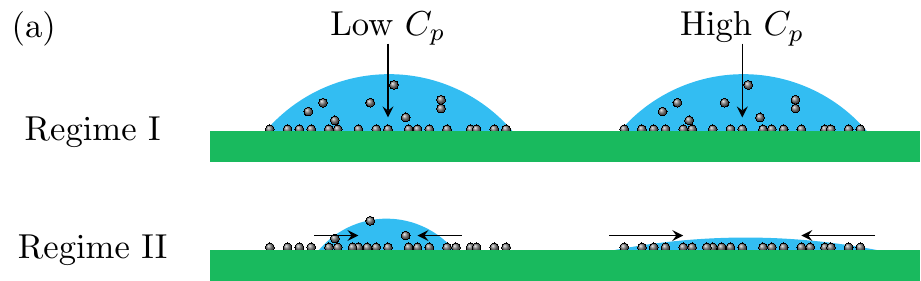}\\
    \includegraphics{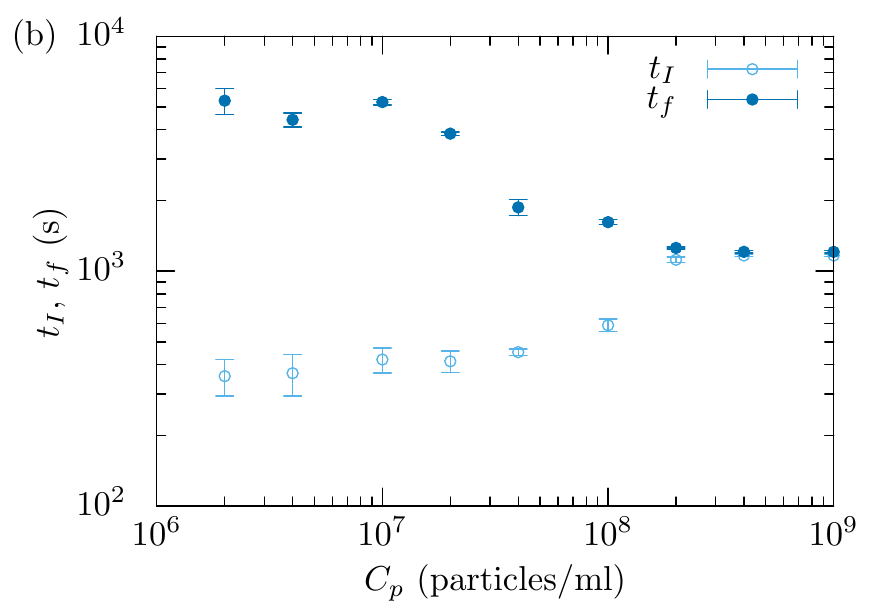}
    \caption{(a) Schematics illustrating the two regimes for low and high particle concentrations.
        (b) Durations of regime I, $t_I$, and the total absorption time, $t_f$, as a function of the particle concentration $C_p$.
        The error bars represent the standard deviation.
    }\label{fig:durations}
\end{figure}

The schematic illustrations in figure \ref{fig:durations}(a) depict the qualitative effect of the particle concentration on the two regimes of drop dynamics.
As the particle concentration increases, particles reinforce the pinning of the contact line already observed on bare hydrogels.
Thus, the duration $t_I$ of regime I increases, which allows more efficient absorption of the drop solvent over a large surface area (figure \ref{fig:durations}(b)).
When the contact line recedes, the area for liquid absorption decreases.
However, for larger particle concentrations, the remaining volume of water in the drop is smaller at $t=t_I$ and a shorter absorption time $t_f$ is observed, as reported also in figure \ref{fig:durations}(b).

For different particle concentrations $C_p$, at low particle concentrations, typically $C_p<1\times 10^7$ particles/ml, $t_I$ and $t_f$ are comparable to the values obtained for a pure water drop of the same volume.
At such low concentrations, the particles weakly affect the transition of regimes I and II.
For a larger concentration, the duration of regime I increases while the total absorption time decreases.
For $C_p> 2\times10^8$ particles/ml, the drop height at $t=t_I$ is comparable to the particle size (figure \ref{fig:durations}(a) and SI Movie).
Therefore, we observe $t_I\approx t_f$.

\subsection{Particle deposition}

As the gel absorbs water, particles are carried toward the gel interface.
In our experiments, we observed that the particles are irreversibly absorbed on the gel, \textit{i.e.} particles are never resuspended once they are adsorbed even when the contact line recedes.
Thus, the final deposition pattern depends on the absorption flux $J(r,t)$ and on the transition between regimes I and II, which drives the wetting area where particles can be deposited \cite{Boulogne2015b}.

\begin{figure}
    \centering
    \includegraphics[width=.9\linewidth]{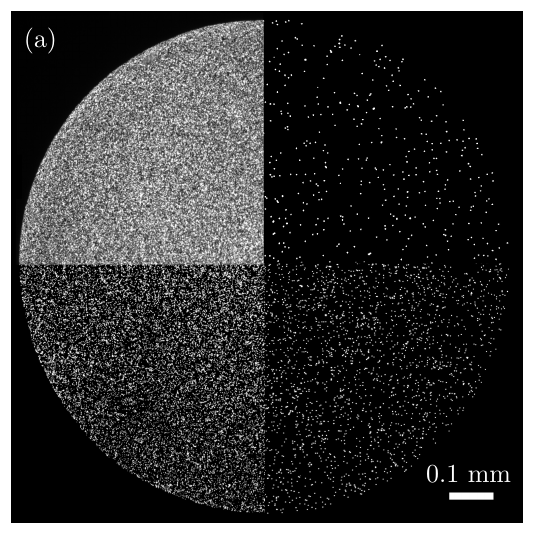}\\
    \includegraphics[width=\linewidth]{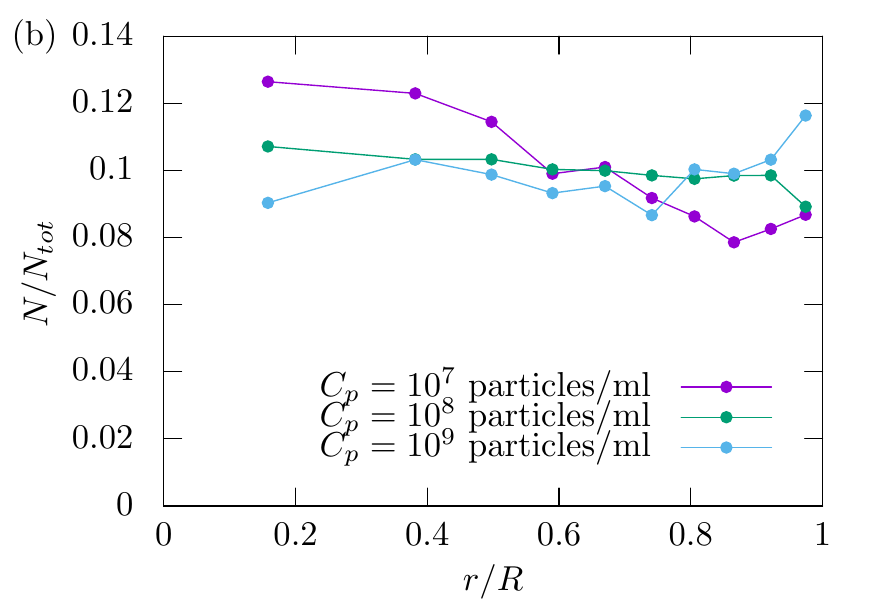}
    \caption{(a) Final deposit for four initial concentrations (from the top right, in the clockwise order): $C_p = 2\times 10^6$, $10^7$, $10^8$ and $10^9$ particles/ml. Images are obtained by fluorescent microscopy.
        (b) Number of particles $N$ non-dimensionalized by the total number $N_{tot}$ as a function of the radial position for three different concentrations.
    }\label{fig:final_profiles}
\end{figure}

The final deposition patterns for four different concentrations ranging from $C_p=2\times 10^6$ to $10^9$ particles/ml are shown in figure \ref{fig:final_profiles}(a).
To quantify the effect of the particle concentration on the particle deposition profile, we used mixtures of non-fluorescent and fluorescent particles.
In the fluorescent mode of the microscope, we detect the individual positions of fluorescent particles.
The blended suspension allows us to apply the same algorithm regarding the particle concentration.
In figure \ref{fig:final_profiles}(b), we show the radial concentration profile obtained from binning the deposit area in concentric rings of the same areas.
For a particle concentration of $C_p=10^{7}$ particles/ml, we recover the profile obtained previously \cite{Boulogne2015b}, which shows a small decay along the radius.
We explained this profile from the combination of the particle deposition that occurs during regimes I and II.
As the particle concentration increases, the trend of absorbed particle number density versus radii is inverted.
A nearly flat profile is observed for $C_p=10^8$ particles/ml and a larger deposition is obtained near the contact line for $C_p=10^9$ particles/ml.
Indeed, the radial profile is similar to the observation made at the end of the regime I for a low particle concentration \cite{Boulogne2015b}.
Thus, for $t_I\approx t_f$, a peak at the drop edge is observed because of a larger absorption flux near the contact line.
However, for lower concentrations, additional particles are deposited when the contact line recedes for $t>t_I$, which allows more particles to deposit in the center.

In the next sections, we rationalize the effect of the particle concentration on the duration of the regime I, namely $t_I$.
In particular, we investigate how the particle concentration affects the contact angle of the liquid on the hydrogel when the contact line depins.
It is particularly challenging to measure the liquid contact angle $\theta_{liq}(t)$ on the hydrogel because the substrate continuously deforms as the absorption proceeds.
Also, direct optical methods are difficult to employ here for two reasons.
First, the contrast of refractive index between the hydrogel and the drop is low, such that the position of the gel-liquid interface cannot be measured by methods based on refraction.
Second, as we study the effect of particles, these particles may interfere with direct visualization.
Thus, we developed another method relying on the presence of particles to measure the gel-liquid interface position and, with further modeling, we obtain the liquid contact angle $\theta_{liq}(t)$ for any particle concentration.

\section{Results and interpretation}

\subsection{Model for the contact angle dynamics}

\begin{figure}
    \centering
    \includegraphics[width=\linewidth]{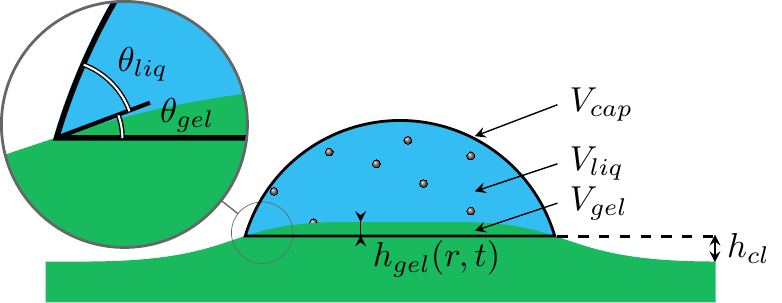}
    \caption{Notations for the calculations including the effect of the swelling of the gel on the liquid contact angle. The volume $V_{gel}$ represents the volume of the swollen gel in the spherical cap drawn in black.}\label{fig:notation_theory}
\end{figure}

We now consider the change in the contact angles with the notations presented in figure \ref{fig:notation_theory}.
We denote $\theta_{cap}$ the apparent contact angle of the liquid drop, which is defined as
\begin{equation}\label{eq:contact_angle_def}
    \theta_{cap}(t) = \theta_{liq}(t) + \theta_{gel}(t),
\end{equation}
where $\theta_{liq}$ is the angle between the gel and the liquid-vapor interface and $\theta_{gel}$ is the angle of the gel at the contact line.

We now consider the time evolution of a spherical cap volume defined as $V_{cap} = V_{liq} + V_{gel}$
where $V_{liq}$ is the liquid volume during the regime I and $V_{gel}$ is the swollen volume of gel beneath the drop (figure \ref{fig:notation_theory}).
The drop has a radius $R$ smaller than the capillary length $\ell_c = \sqrt{\gamma/(\rho_\ell g)}\approx 2.7$ mm, where $\gamma$ is the liquid-vapor surface tension, $\rho_\ell$ is the liquid density and $g$ is the gravity.
Thus, the drop shape is approximated as a spherical cap, and for small angles, we have
\begin{equation}\label{eq:Vcap}
    V_{cap} = \frac{\pi R^3 \theta_{cap}}{4} = V_{liq} + V_{gel}.
\end{equation}

During regime I, we have shown previously that the flux $J(r,t)$ of water from the drop to the gel is spatially homogeneous in the center of the drop \cite{Boulogne2015b}.
The flux is larger near the contact line but this deviation is localized to $1/10$ of the drop radius \cite{Boulogne2015b}.
Therefore, we neglect this deviation and we consider the time evolution of the flux as
\begin{equation}
    J(t) \approx \frac{ \sqrt{{\cal D}} \delta }{ \sqrt{t}},
\end{equation}
where ${\cal D}=250$ $\mu$m$^2$/s is an effective diffusion coefficient and $\delta\approx 0.18$ is the swelling ratio.
The swelling ratio is defined as $\delta = \frac{V_{sw}^{w}}{V_{sw}^{tot}}  - \frac{V_i^{w}}{V_{i}^{tot}}$, where $V_i^{w}$ and $V_{sw}^{w}$ are, respectively, the volumes of water in the initial and swollen states, and $V_{i}^{tot}$ and $V_{sw}^{tot}$ are, respectively, the volumes of gel at the initial and swollen states.
Thus, the time evolution of the liquid volume $V_{liq}$ during this regime is
\begin{equation}\label{eq:VI}
    V_{liq}(t) \approx V_0 - 2 \delta  S \sqrt{{\cal D} t},
\end{equation}
where $V_0$ is the initial volume and $S=\pi R^2$ is the wetting area.
By combining equations (\ref{eq:Vcap}) and (\ref{eq:VI}) into equation (\ref{eq:contact_angle_def}), we obtain
\begin{equation}\label{eq:contact_angle_result}
    \theta_{liq}(t) \approx \frac{4}{\pi R^3}\left(V_0 - 2\pi R^2 \delta  \sqrt{{\cal D}t} + V_{gel}(t) \right) - \theta_{gel}(t).
\end{equation}
Also, we remark that neglecting the swelling of the substrate is equivalent to set $V_{gel}(t)=0$ and $\theta_{gel}(t)=0$ in equation (\ref{eq:contact_angle_result}), \textit{i.e.}

\begin{equation}\label{eq:contact_angle_noswelling}
    \theta_{liq}(t) \approx \frac{4}{\pi R^3}\left(V_0 - 2\pi R^2 \delta  \sqrt{{\cal D}t} \right).
\end{equation}

By inspection of equations (\ref{eq:contact_angle_result}) and (\ref{eq:contact_angle_noswelling}), we can see that the swelling correction $V_{gel}(t) - \frac{\pi R^3}{4} \theta_{gel}(t)$ included in equation (\ref{eq:contact_angle_result}) is the difference between the actual shape of the gel and a spherical cap with the gels angle at the contact line.
In the next section, we present the measurements to obtain $\theta_{gel}(t)$ and $V_{gel}(t)$ to compute $\theta_{liq}(t)$ from equation (\ref{eq:contact_angle_result}).

\subsection{Measurement of $\theta_{liq}(t)$}

To estimate the quantities $V_{gel}(t)$ and $\theta_{gel}(t)$, we first measure the time evolution of the gel-liquid interface position $h_{gel}(r,t)$; see figure \ref{fig:notation_theory}.
To describe the depinning transition, we assume that the absorption dynamics is independent of the particle concentration.
This is a reasonable assumption since the highest concentration of particles leads to a deposit that covers around $\approx 20$\% of the surface.
This deposit has a permeability very high compare to the one of the hydrogel.
Therefore, the particles cannot apply a significant resistance to the flow and can be neglected when considering the absorption dynamic.
Thus, the measurement of $\theta_{liq}(t)$ is also independent of $C_p$.
Since $t_I$ increases with $C_p$, the measurement of $\theta_{liq}(t)$ for the largest concentration will provide $\theta_{liq}(t_I)$.

To measure $h_{gel}(r,t)$ (see figure \ref{fig:notation_theory}), we use a suspension at a concentration $C_p=1\times 10^{9}$ particles/ml.
As some particles are rapidly adsorbed at the gel interface, we use their fluorescence to detect the position of the interface.

With a confocal microscope, we take images in a vertical plane perpendicular to the gel surface by scanning along the $z$ direction with steps of $2.7$ $\mu$m.
Images are taken along the drop radius with a resolution of 512 pixels and with a width of 50 pixels.
At this magnification, 512 pixels represent 1.5 mm (\textit{i.e.} a resolution of 3$\mu$m/px), which is larger than the drop radius $R\approx 1.1$ mm.
The $z$-scan is repeated every 30 s.

\begin{figure}
    \centering
    \includegraphics{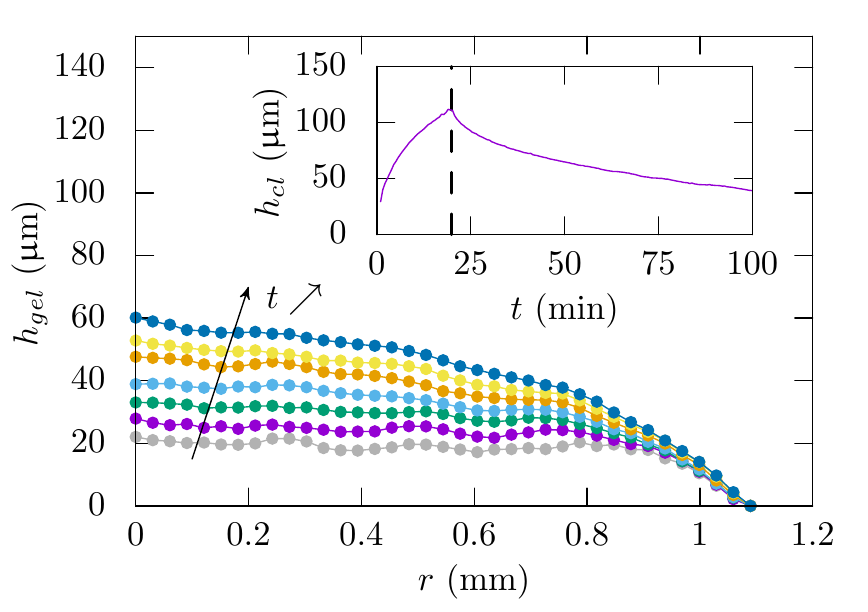}
    \caption{Time evolution of the gel interface position $h_{gel}(r,t)$.
        The reference is the vertical position of the contact line $h_{cl}(R,t)$, which is set to zero.
        Times range from $t=2$ min to $t=20$ min, with a time increment of 3 min.
        The inset shows the time evolution of the contact line vertical position $h_{cl}(t)$.
        The vertical dashed line represents $t=t_I\approx t_f \approx 20$ min.
    }\label{fig:gel_shape}
\end{figure}

For each picture, the sum of pixel intensities is calculated in a sliding windows of ($30\times50$) pixels moving along the radius with a step of $10$ pixels ($\approx 30$ $\mu$m).
Thus, we obtain an intensity $i(r,z,t)$ for which the radial position corresponds to the center of the sliding windows.
For a given radial position $r_i$ and time $t_i$, the intensity $i(r_i, z, t_i)$ presents a maximum located at the gel-liquid interface due to fluorescent particle adsorption.
The position of the maximum is determined from a fit with a Lorentzian curve to achieve a subpixel precision.
As a result, we obtain the interface position of the gel $h_{gel}(r,t)$ reported in figure \ref{fig:gel_shape}.

\begin{figure}
    \centering
    \includegraphics{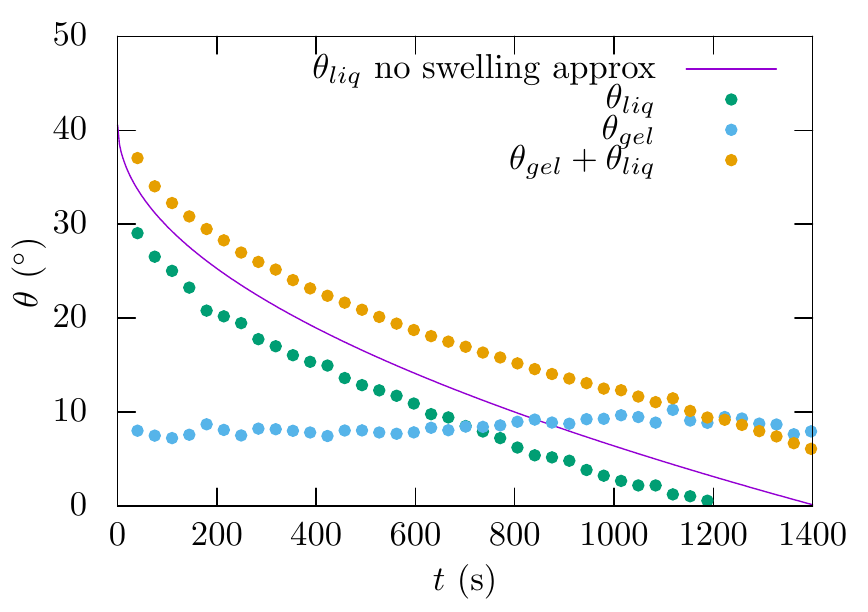}
    \caption{Time evolution of the liquid contact angle $\theta_{liq}$ on the gel  for data shown in figure \ref{fig:gel_shape}.
        The solid line represents equation (\ref{eq:contact_angle_noswelling}) where the swelling of the gel is neglected and the green points are given by equation (\ref{eq:contact_angle_result}) accounting for the gel swelling.
        Blue points represent the time evolution of the gel angle $\theta_{gel}$ at the contact line and the orange points, the sum $\theta_{gel}+\theta_{liq}$.
    }\label{fig:liquid_angle}
\end{figure}

From figure \ref{fig:gel_shape}, we use a trapezoidal method to calculate the volume $V_{gel}(t) = 2\,\pi \int_0^R h_{gel}(r,t)\, r\,{\rm d} r$ and we also calculate the gel angle $\theta_{gel}(t)$ from the tangent of $h_{gel}(r,t)$ at the contact line.
These values are reported in figure \ref{fig:liquid_angle} and shows only a small variation during our observations.
The first minute of swelling is not available due to the time resolution of our experiment.

Then, with equation (\ref{eq:contact_angle_result}), we estimate $\theta_{liq}(t)$, which is given in figure \ref{fig:liquid_angle}.
We also plot in figure \ref{fig:liquid_angle}, the estimate of $\theta_{liq}(t)$ from equation (\ref{eq:contact_angle_noswelling}), which does that not account for the swelling effect.
For a given volume of liquid, \textit{i.e.} a given time (see equation (\ref{eq:VI})), the contact angle is smaller due to the inclined position of the gel interface.
Nevertheless, both curves have a similar trend and the correction coming from the swelling of the hydrogel remains small.

\subsection{Effect of the particle concentration on the contact line depinning}

In the literature on contact line hysteresis, it is common to define the hysteresis $H$ as the difference between the advancing and the receding force that has been applied on a contact line to trigger its motion.
A clever measurement of this hysteresis has been provided by di Meglio and Qu\'er\'e for contact line motion on fibers \cite{Meglio1990}.
For a surface covered with dilute defects, is has been shown theoretically \cite{Joanny1984,Pomeau1985} and experimentally \cite{Marsh1993} that the hysteresis is proportional to the density of defects.
For larger defect surface concentration $C_s$, numerical and experimental studies \cite{Meglio1992,Crassous1994} indicate that the hysteresis grows as $C_s^{0.7}$.

\begin{figure}
    \centering
    \includegraphics{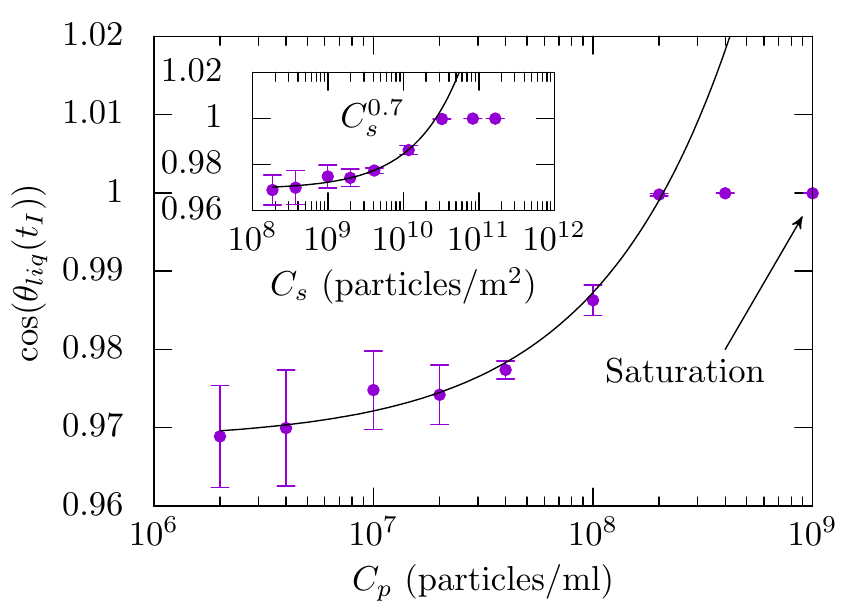}
    \caption{
        The points represent $\cos(\theta_{liq}(t_I))$ as a function of the initial particle concentration $C_p$ for 1 $\mu$m diameter particles.
        The error bars represent the uncertainty reported from data presented in Figure \ref{fig:durations}(b).
        The solid line is a guide for the eye.
        The inset shows $\cos(\theta_{liq}(t_I))$ as a function of the defect surface concentration $C_s$ calculated from equation (\ref{eq:surf_conc}).
        The solid line represents the equation $\cos(\theta_{liq}(t_I)) \propto C_s^{0.7}$.
        }\label{fig:hysteresis}
\end{figure}

As we established the evolution of the liquid contact angle $\theta_{liq}(t)$ on the hydrogel (figure \ref{fig:liquid_angle}), we can interpret the durations of regime I that depends on the particle concentration (figure \ref{fig:durations}(b)), as a critical receding contact angle.
Thus, using $t_I (C_p)$ from figure \ref{fig:durations}(b) and $\theta_{liq}(t_I)$ from figure \ref{fig:liquid_angle}, we derive $\cos(\theta_{liq}(t_I))$ at the receding time as a function of the particle concentration $C_p$ as represented in figure \ref{fig:hysteresis}.

For particle concentrations typically larger than $5\times 10^7$ particles/ml, the direct particle identification and counting is particularly difficult.
To estimate the particle surface concentration  at time $t_I$ for all experiments, we consider the absorbed volume of the drop $2\delta S \sqrt{{\cal D} t_I}$ \cite{Boulogne2015b}.
Thus, the particle surface concentration  adsorbed at the surface at $t=t_I$ is

\begin{equation}\label{eq:surf_conc}
C_s = 2 \delta \sqrt{{\cal D} t_I} C_p.
\end{equation}

Our data shows that the hysteresis increases with the particle concentration and then saturates for $C_p>2\times 10^8$ particles/ml or for an average coverage of $C_s\approx 3\times 10^{10}$ particles/m$^2$.
At this concentration, the particles apply a sufficient force to maintain pinned the contact line during the entire absorption process.
Furthermore, as suggested by the experimental literature on solid surfaces, a power-law $\cos(\theta_{liq}(t_I)) \propto C_s^{0.7}$ also successfully describes our results.

Our observations are in agreement with the results obtained by Kajiya \textit{et al.} \cite{Kajiya2011} for pure liquid water drops.
For very dilute suspensions, the critical liquid contact angle below which the contact line recedes is small.
Nevertheless, this angle is not zero and for our gels is $\theta_{liq} \approx 14^\circ$.
The results presented in figure \ref{fig:hysteresis} shows that the presence of micron size particles at the contact line allows us to reduce continuously this critical angle to zero.

\section{Conclusion}

In this paper, we studied the effect of the particle concentration on the dynamics of a drop absorbed by a swelling hydrogel.
The dynamics of a pure water drop is composed of a first regime where the contact line is pinned on the hydrogel and a second regime where the contact line recedes.
As the particle concentration increases, the duration of regime I increases.
Since the absorption area is larger during this first regime, the total absorption time decreases with the particle concentration.
For micron size particles, we observed that above a surface concentration of $2\times 10^{10}$ particles/m$^2$, the contact line remains pinned throughout the entire absorption process.
To quantify the critical receding contact angle $\theta_{liq}(t_I)$, we developed a method based on the detection of the gel-water interface position.
We have shown that $\cos(\theta_{liq}(t_I)) \propto C_s^{0.7}$, in agreement with the literature related to the contact line hysteresis on solid substrates containing defects \cite{Meglio1992,Crassous1994} .

The results reported here illustrate that the addition of particles on the surface of a hydrogel modifies the receding contact angle.
In the situation of a colloidal drop absorbed by the hydrogel, we have shown that the particle surface concentration profile is modified by changing the initial particle concentration. 
We related this effect to the variation of the receding contact angle due to the  particles adsorbed during the swelling process.
In particular, we found for our system that a concentration of $C_s\approx 3\times 10^{10}$ particles/ml leads to an homogeneous coating, while a lower concentration generates a larger deposition in the center of the deposit.

These results extend our previous findings on the  homogeneous coatings of particles on hydrogels \cite{Boulogne2015b}, which can be useful for the realization of functionalized interfaces of hydrogels.
In terms of applications, these questions are particularly crucial for biomedical applications such as soft contact lenses \cite{Holly1975,Ketelson2005}, functional biogels \cite{Peppas2000,Seliktar1124} or tissue engineering \cite{Rose2014}.
More generally, we believe that adsorption of particles can inspire future studies to tune the wetting properties of hydrogels.
From the fundamental point of view, the field of solid capillarity raises important questions on the wetting of soft materials, the role of soft solid interfaces and the importance of physical and chemical properties \cite{Andreotti2016}. 

Future work will be devoted to the importance of chemical and physical properties of the particles on the final wetting properties.
Indeed, we expect that the size of the defects as well as the hydrophilicity of the chemical groups grafted on the particles are significant.
Since the experiments described in this paper allows measurement of the pinning force caused by the defects, it is possible to compare this force for different particle sizes or different particle surface chemistry.
These measurements might be useful to predict how a deposit of particles changes the spreading of water on a hydrogel surface.

\section{Appendix}

\begin{figure}
    \centering
    \includegraphics{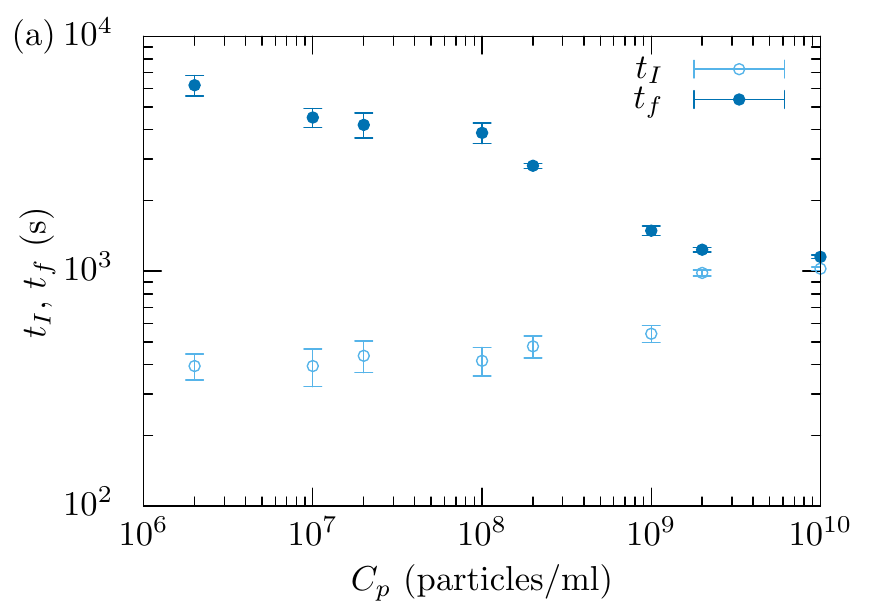}\\
    \includegraphics{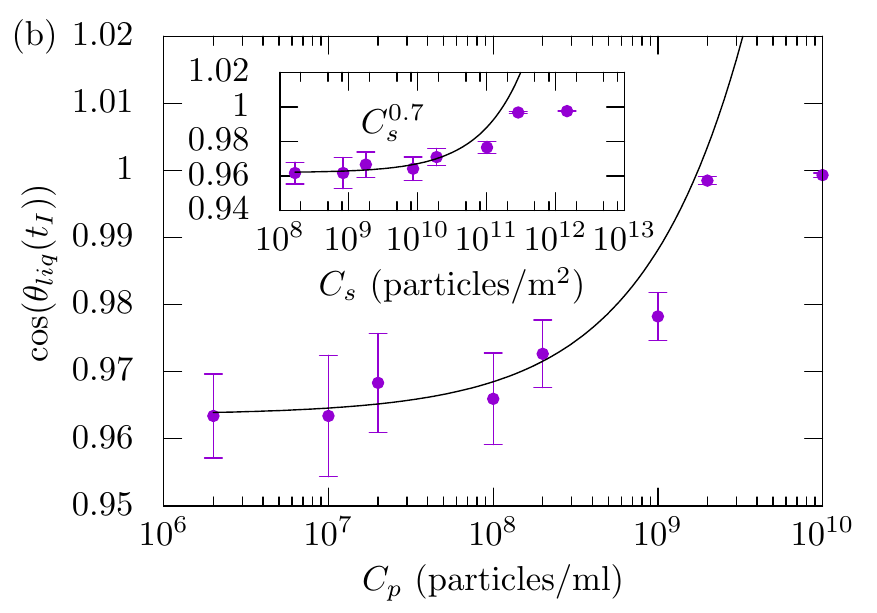}
    \caption{ Data for $0.2$ $\mu$m diameter particles.
        (a) Durations of regime I, $t_I$, and of the total absorption time $t_f$ as a function of the particle concentration $C_p$.
        (b) The points represents $\cos(\theta_{liq}(t_I))$ as a function of the initial particle concentration $C_p$.
        The solid line is a guide for the eye.
        The inset shows $\cos(\theta_{liq}(t_I))$ as a function of the defect surface concentration $C_s$ calculated from equation (\ref{eq:surf_conc}).
        The solid line represents the equation $\cos(\theta_{liq}(t_I)) \propto C_s^{0.7}$.
    }\label{fig:annex02um}
\end{figure}

In addition to the results presented in the main text, which are obtained for $1$ $\mu$m diameter particles, we reproduced the experiments for $0.2$ $\mu$m diameter particles.
Figure \ref{fig:annex02um} shows equivalent data to figures \ref{fig:durations}(b) and \ref{fig:hysteresis}.
Both plots show trends similar to the observations made for $1$ $\mu$m diameter particles.
However, the particle concentration must be increased by a factor $\approx 10$ to obtain $t_I\approx t_f$.

\section{Acknowledgements}
F.B. acknowledges that the research leading to these results received funding from the People Programme (Marie Curie Actions) of the European Union's Seventh Framework Programme (FP7/2007-2013) under REA grant agreement 623541.

\bibliography{article_pinning}
\bibliographystyle{unsrt}


\end{document}